\documentclass[prb,aps,twocolumn,floats]{revtex4}
\usepackage{graphicx}
\input{psfig}

\begin{document}

\title{Pressure-induced changes in the optical properties of quasi-one-dimensional 
$\beta$-Na$_{0.33}$V$_2$O$_5$}

\author{Simone Frank and Christine A. Kuntscher$^*$}
\address{Lehrstuhl f\"{u}r Experimentalphysik\ II, Universit\"at Augsburg, D-86135 Augsburg, Germany}
\author{Ivan Gregora}
\address{Institute of Physics, Academy of Sciences of the Czech Republic, Na Slovance 2, 182 21 Prague 8, Czech Republic}
\author{Touru Yamauchi and Yutaka Ueda}
\address{Institute for Solid State Physics, University of Tokyo, 5-1-5 Kashiwanoha, Kashiwa, Chiba 277-8581, Japan}

\date{\today}

\begin{abstract}
The pressure-induced changes in the optical properties of $\beta$-Na$_{0.33}$V$_2$O$_5$ single crystals
at room temperature were studied by polarization-dependent Raman and far-infrared reflectivity
measurements under high pressure.
From the changes in the Raman- and infrared-active phonon modes in the pressure range
9 - 12~GPa a transfer of charge between the different V sites can
be inferred. The importance of electron-phonon coupling in the low-pressure regime
is discussed.
\end{abstract}


\maketitle
\section{Introduction}

\begin{figure}[t]
\centerline{\psfig{file=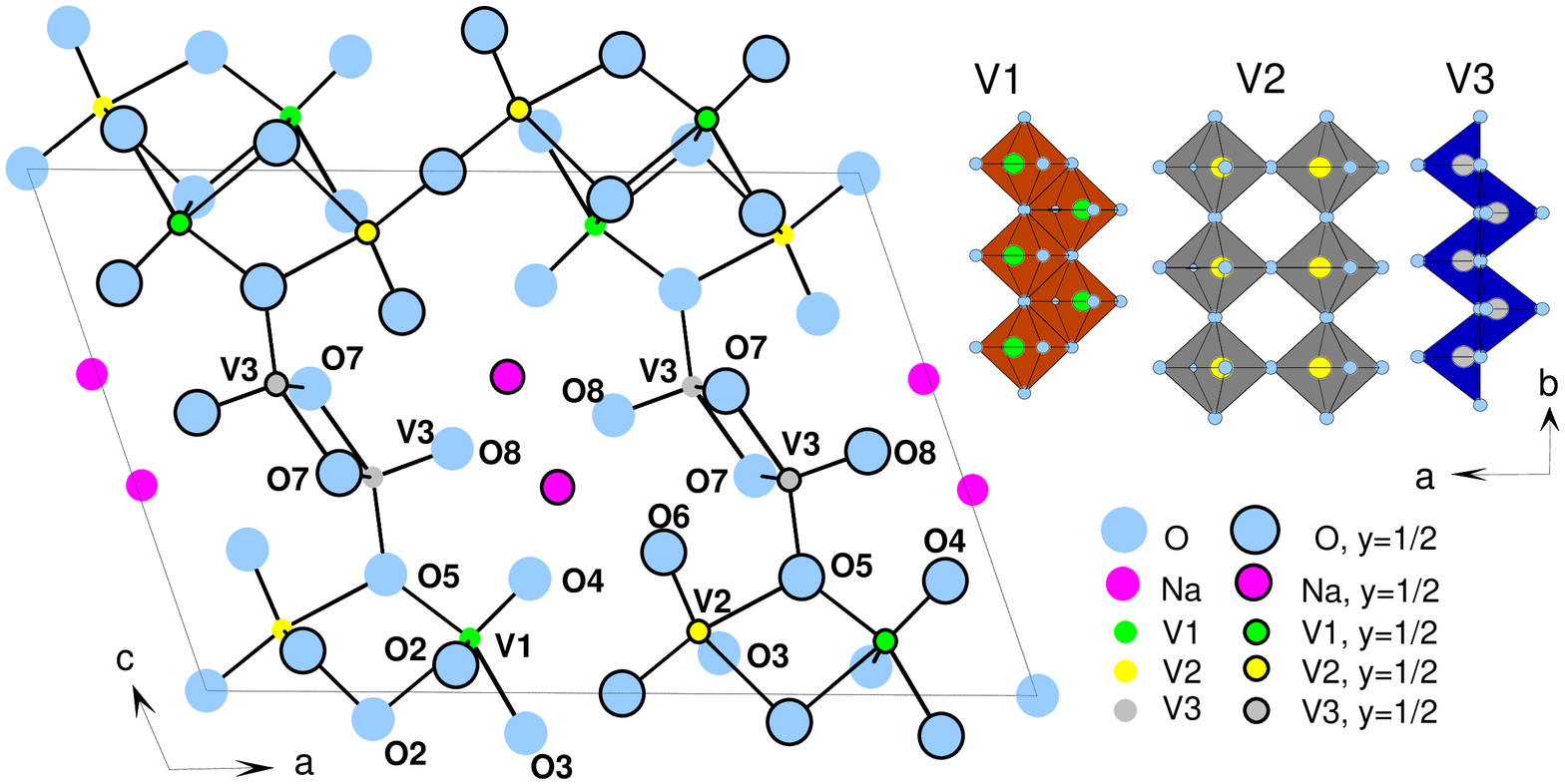,width=85mm,clip=}}
\caption{(Color online) Crystal structure of $\beta$-Na$_{0.33}$V$_2$O$_5$
projected in the (010) plane.\cite{Wadsley55} The vanadium oxide
two-leg ladder as well as the two kinds of vanadium oxide chains
are also shown.} \label{struc}
\end{figure}

In solid state matter with complex crystal structure the
distribution of charge on the various structural entities, like
chains or ladders of polyhedra, generally is one key for understanding
the observed electronic and magnetic properties. In the presence
of strong electronic correlations, especially in systems with
reduced dimensions, these systems often exhibit an
ordering of the charge carriers or a charge disproportionation.
In some rare cases superconductivity occurs in direct
vicinity to the charge-ordered phase. The low-dimensional vanadate
$\beta$-Na$_{0.33}$V$_2$O$_5$ is such an example: It shows a
superconduc\-ting phase in the pressure range between 7~GPa to 9~GPa
below $\approx$9~K.\cite{Yamauchi02} For lower pressures a phase
of charge ordering/charge disproportionation is found.\cite{Yamauchi02,Itoh06}

$\beta$-Na$_{0.33}$V$_2$O$_5$ crystalizes in a monoclinic
tunnel-like structure built by three kinds of chains along the
\emph{b} axis which consist of three inequivalent vanadium sites.
Along the \emph{b} axis the edge-shared (V1)O$_6$ octahedra form a
zigzag double chain. The (V2)O$_6$ octahedra form a two-leg ladder
by corner sharing and the (V3)O$_5$ polyhedra form zigzag double
chains. There are two possible sodium sites located in the tunnels
along the \emph{b} axis. They can be represented as a two-leg
ladder along the \emph{b} axis. For a sodium stoichiometry of 0.33
half of the sites is occupied. At room temperature the sodium
atoms are statistically distributed over these
sites.\cite{Wadsley55} In Fig.\ \ref{struc} a projection of the
crystal structure, the two types of chains, and the two-leg ladder
formed by vanadium oxide polyhedra and octahedra are shown. Recent NMR
studies in combination with theoretical investigations suggest
that regarding the dominant electronic interactions the system
is rather to be described as consisting of weakly coupled V2-V2
and V1-V3 ladders.\cite{Doublet05,Itoh06}

Dc resistivity measurements on $\beta$-Na$_{0.33}$V$_2$O$_5$
reveal a one-dimensional metallic character, with the lowest
resistivity along the \emph{b} axis, i.e., along the chain
direction.\cite{Yamada99,Ueda01,Heinrich04} This is in agreement with the
optical properties of the material:\cite{Presura03} For the
polarization of the incident radiation along the chain direction,
\textbf{E}$\parallel$\emph{b}, the reflectivity is high and a
Drude contribution is found at room temperature, whereas in the perpendicular
direction an insulating behavior is observed.

Up to now the nature of the pressure-induced superconductivity and
the role of the distribution of charge for the superconducting
phase are not clear. A first pressure-dependent study of the
infrared reflectivity at room temperature revealed the electronic
and lattice dynamical properties of $\beta$-Na$_{0.33}$V$_2$O$_5$
single crystals along different
directions.\cite{Kuntscher05,Kuntscher06,Kuntscher062} The results
of this study suggest the possible role of polaronic
quasiparticles for the superconductivity: For \textbf{E}$\parallel$\emph{b}
a pronounced mid-infrared is observed, which was claimed\cite{Presura03} to 
be of small-polaron origin. For pressures up to 12~GPa, this mid-infrared band
shifts to smaller frequencies with increasing pressure, in agreement with
small-polaron theory.\cite{Kuntscher05,Kuntscher06,Kuntscher062}
Above 12~GPa this trend is reversed. Further significant spectral changes
are induced at around 12 GPa, like the development of additional excitations.
It was suggested that the additional excitations are related to a
redistribution of charge among the different V sites.
Based on the infrared data
a pressure-induced structural phase transition as an explanation
for the additional excitations appeared unlikely but could not
be ruled out.\cite{Kuntscher05}

Raman spectroscopy, as a complementary tool to infrared spectroscopy, can
give additional important information on the lattice dynamical properties.
We therefore carried out a polarization-dependent Raman study on
$\beta$-Na$_{0.33}$V$_2$O$_5$ single crystals under high pressure.
We find significant changes in the Raman modes induced for pressures
9 - 12~GPa, in agreement with the pressure-dependent frequency positions of
the far-infrared phonon modes. Several scenarios are considered as possible
explanations for these findings, like structural phase transition,
amorphization or charge redistribution.
Furthermore, the possible relevance of electron-phonon coupling in
$\beta$-Na$_{0.33}$V$_2$O$_5$ is discussed.

\section{Experiment}
\label{sectionexperiment}

The investigated $\beta$-Na$_{0.33}$V$_2$O$_5$ single crystals
were grown according to Ref.\ \onlinecite{Yamada99}. The sample
quality was checked by infrared and dc resistivity measurements.
Polarization-dependent Raman measurements at room temperature were
carried out in backscattering geometry with a
Renishaw System 1000 Micro-Raman Spectrometer equipped with a
notch filter (130~cm$^{-1}$ cutoff frequency) and a CCD multichannel
detector. The 632.8~nm He-Ne laser line was used for excitation.
To focus the beam on the sample an Olympus objective with a 20x
magnification and a 21~mm working distance was attached to the
microscope. The spot size on the sample was 5~$\mu$m. The studied
frequency range extends from 140~cm$^{-1}$ to 1200~cm$^{-1}$. For
the high pressure experiment a diamond anvil cell equipped with
type IIA diamonds was used. A small sample of the size
of 80x80~$\mu$m$^2$ was cut and placed in the hole (150~$\mu$m) of
a steel gasket. Finely ground KCl powder served as quasi-hydrostatic
pressure transmitting medium. A small ruby chip was added to
determine the pressure with the ruby luminescence
method.\cite{Mao86} The reproducibility of the results was checked
by two experimental runs.

In addition, we carried out polarization-dependent reflectivity measurements
in the far-infrared range (200 - 650~cm$^{-1}$) at the infrared beamline
of the synchrotron radiation source ANKA. A Bruker IFS 66v/S FT-IR spectrometer
in combination with an infrared microscope (Bruker IRscope II), equipped
with a 15$\times$ magnification objective was employed.
For pressure generation up to 14~GPa a diamond anvil cell with type IIA diamonds
and KCl powder as pressure medium were used.
Reflectivity spectra $R_{s-d}$ were measured at the interface between the
sample and diamond anvil. Spectra taken at the inner diamond-air
interface of the empty cell served as the reference for normalization of the
sample spectra. Schemes for the geometries of the sample and reference
measurements are given in Ref.\ \onlinecite{Kuntscher05}.
Variations in source intensity were taken into account by
applying additional normalization procedures.
To obtain the frequency positions of the phonon modes,
the reflectivity spectra were fitted with the Drude-Lorentz model
combined with the normal-incidence Fresnel equation, taking into account the
known refractive index of diamond.

\begin{figure}[t]
\centerline{\psfig{file=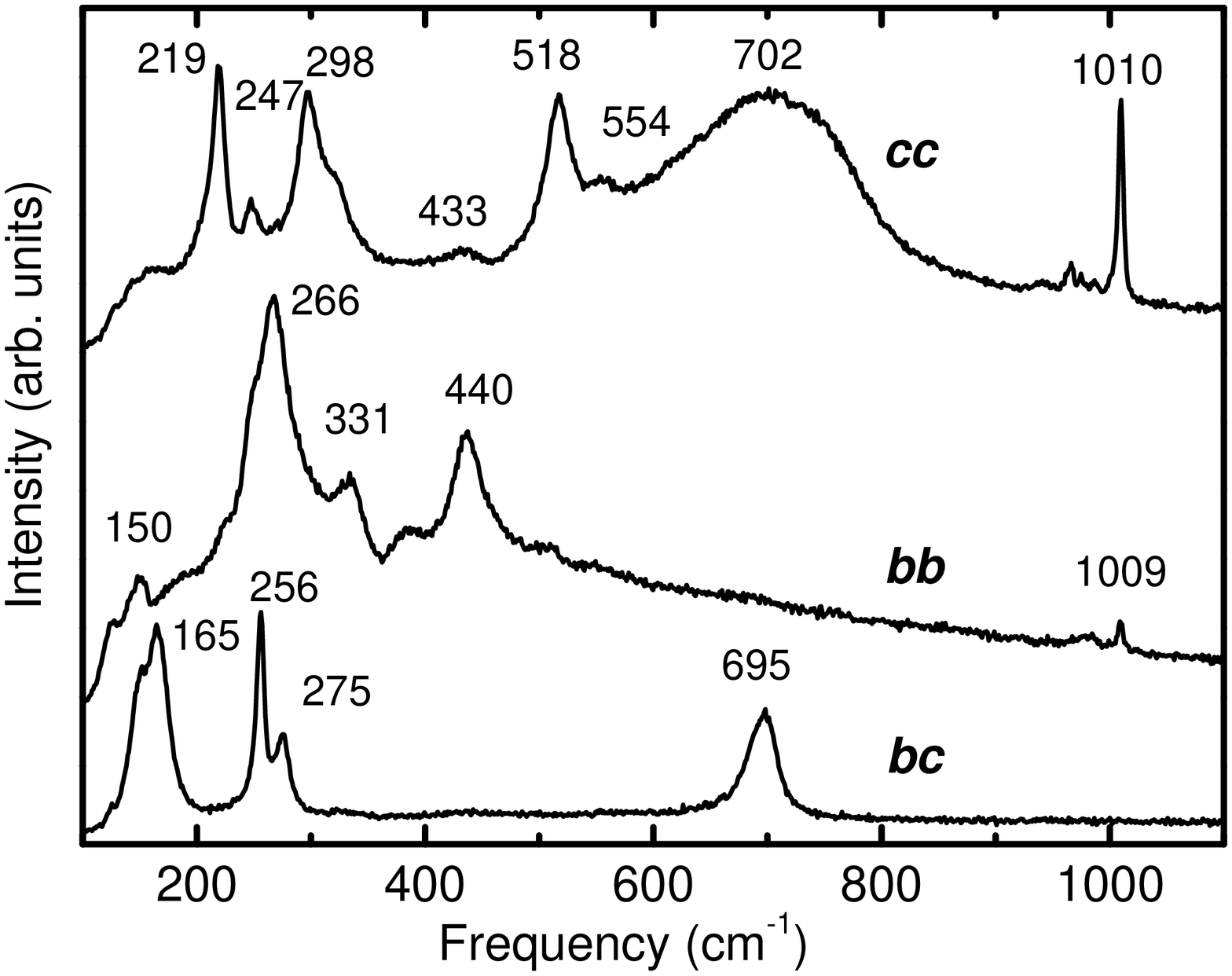,width=80mm,clip=}}
\caption{Room-temperature Raman spectra of $\beta$-Na$_{0.33}$V$_2$O$_5$ at
ambient pressure for the polarizations \emph{cc}, \emph{bb}, and \emph{bc}.
The spectra are shifted along the vertical axis for clarity.} \label{freespec}
\end{figure}

\section{Results}
\label{sectionresults}

\subsection{Ambient-pressure Raman spectra}
In Fig.\ \ref{freespec} the room-temperature ambient-pressure Raman spectra
of $\beta$-Na$_{0.33}$V$_2$O$_5$ are shown for parallel and crossed polarizations. The
spectra are shifted along the vertical axis for clarity. For
\emph{cc} polarization seven modes (219, 247, 298, 433, 518, 554, 1010~cm$^{-1}$)
and one broad peak-like structure at 702~cm$^{-1}$ are observed. For \emph{bb}
polarization five strong modes (150, 266, 331, 390, 440, 1009~cm$^{-1}$) are found.
The crossed \emph{bc} polarization shows five modes (154, 165, 256, 275, 695~cm$^{-1}$)
in the measured range. In Table\ \ref{Rphonon} we list the different modes together
with their assignments. The mode assignment is based on a comparison with Raman data
of the closely related compound $\beta$-Ca$_{0.33}$V$_2$O$_5$:\cite{Popovic03} Both
$\beta$-Ca$_{0.33}$V$_2$O$_5$ and $\beta$-Na$_{0.33}$V$_2$O$_5$
crystallize in the same monoclinic crystal structure, with a doubling of the
unit cell in $\beta$-Ca$_{0.33}$V$_2$O$_5$ along the \emph{b} direction due to an
ordering of the Ca atoms (similar to the low-temperature phase of
$\beta$-Na$_{0.33}$V$_2$O$_5$, see Refs.\ \onlinecite{Isobe00,Yamura02}). Therefore, it
is possible to compare the ambient-pressure Raman data of
$\beta$-Na$_{0.33}$V$_2$O$_5$ with those of
$\beta$-Ca$_{0.33}$V$_2$O$_5$.\cite{Popovic03,Isobe00,Yamura02}
The phonon modes below 500~cm$^{-1}$ originate from the bond bending
vibrations, while those at higher frequencies can be assigned to the
stretching vibrations of the V-O polyhedra and octahedra.
The highest-frequency modes at 1010 and 1009 \nolinebreak cm$^{-1}$ are
attributed to the V3-O8 and V1-O4 bond stretching vibrations, respectively,
since these are the shortest bonds.

The broad peak-like structure at 702~cm$^{-1}$ for \emph{cc} polarization
is also observed in the Raman spectra of several closely related vanadate compounds, the
most prominent one being the $\alpha$'-NaV$_2$O$_5$. $\alpha$'-NaV$_2$O$_5$ was extensively
studied because of its interesting low-temperature properties, namely
a spin-Peierls-like phase transition at $T_c$=35 K with the simultaneous
occurrence of charge disproportionation, lattice dimerization, and
spin-gap formation.\cite{Isobe98}
Despite the differences in the crystal structure of $\alpha$'-NaV$_2$O$_5$ compared
to that of $\beta$-Na$_{0.33}$V$_2$O$_5$ [and other members of the family
$\beta$-$A$$_{0.33}$V$_2$O$_5$ ($A$=Sr,Ca,Na...)] \cite{Popovic03}
-- the former being two-dimensional, the latter one-dimensional in nature --
there are remarkable similarities in their Raman and optical conductivity spectra.
These similarites are due to the fact that the electronic structures
of the two compounds are based on similar structural units, as was
shown by recent extended H\"{u}ckel tight-binding calculations.\cite{Doublet05}

In $\alpha$'-NaV$_2$O$_5$ a similar broad Raman mode is found at
around 650~cm$^{-1}$ in \emph{aa}
polarization.\cite{Konstantinovic02,Golubchik97,Bacsa00,Fischer99,Konstantinovic99,Loa99}
Several scenarios have been proposed to explain 
this broad Gaussian-like band: (i) electric dipole transitions
between the crystal field split V $3d$ levels, (ii) magnon
scattering, and (iii) a mode due to strong electron-phonon coupling.
Up to now, a consensus has not been reached on this issue. On the
other hand, there are additional hints for the relevance of
electron-phonon coupling in $\alpha$'-NaV$_2$O$_5$: the phonon
modes close to the broad band are asymmetric and could be fitted
with a Fano profile.\cite{Konstantinovic02,Popovic02,Loa99}
Generally, such a mode asymmetry is assigned to the interaction
between a discrete state (lattice vibration) and an electronic
continuum. In coparison, in our ambient-pressure {\it cc} Raman
spectrum of $\beta$-Na$_{0.33}$V$_2$O$_5$ an asymmetric profile of
the mode located at 518~cm$^{-1}$, close to the broad band at
around 702~cm$^{-1}$, is not obvious.

\begin{figure}[t]
\centerline{\psfig{file=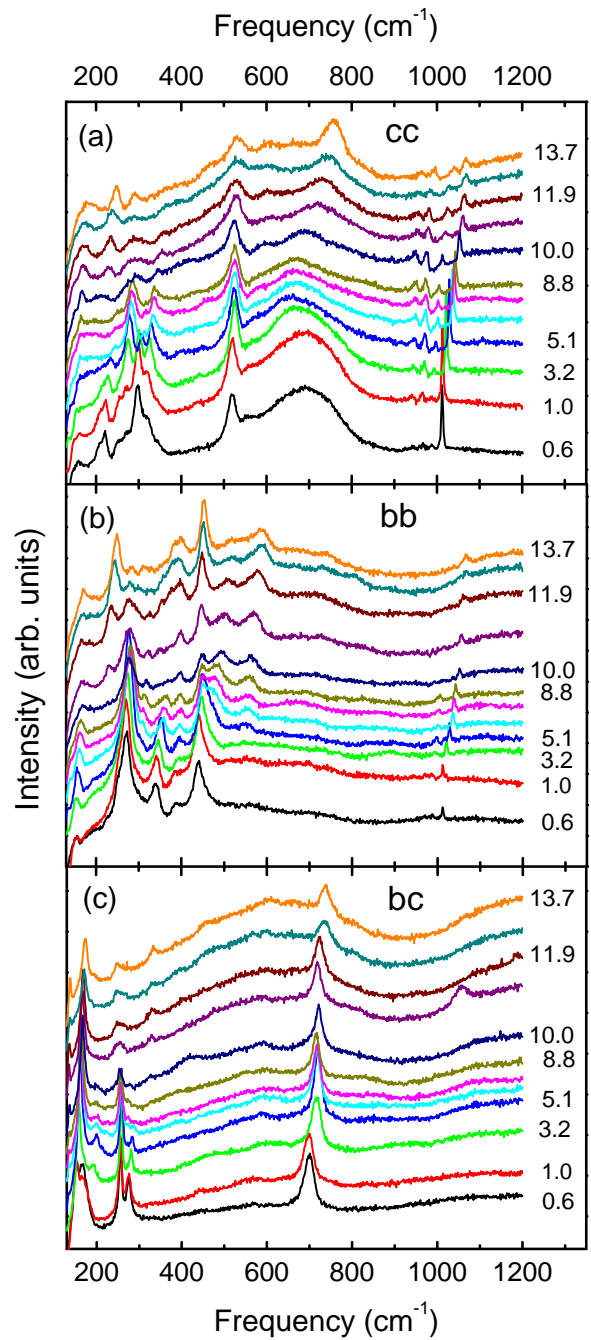,width=80mm,clip=}}
\caption{(Color online) Pressure-dependent Raman spectra of
$\beta$-Na$_{0.33}$V$_2$O$_5$ at room temperature for (a) \emph{cc},
(b) \emph{bb}, and (c) \emph{bc} polarization. The spectra are plotted
with an offset for clarity. The numbers give the applied pressures in GPa.}
 \label{pressall}
\end{figure}

\subsection{Pressure-dependent Raman and far-infrared reflectivity spectra}

The results of the pressure-dependent Raman measurements 
are shown in Fig.\ \ref{pressall}. The spectra are shifted along
the vertical axis for clarity. Due to the diamond absorption the
intensity is reduced and therefore some of the ambient pressure
modes are not detectable.

One important finding is that the Raman spectra do not change
fundamentally in the whole studied pressure range. Most of the
spectral features are present up to the highest applied pressure.
The observed changes are continuous, and mainly consist of a weakening or broadening of
the spectral features. Therefore, we can rule out an amorphization of the sample, 
in agreement with earlier conclusions based on our pressure-dependent mid-infrared 
data.\cite{Kuntscher05} 

\begin{figure}[t]
\centerline{\psfig{file=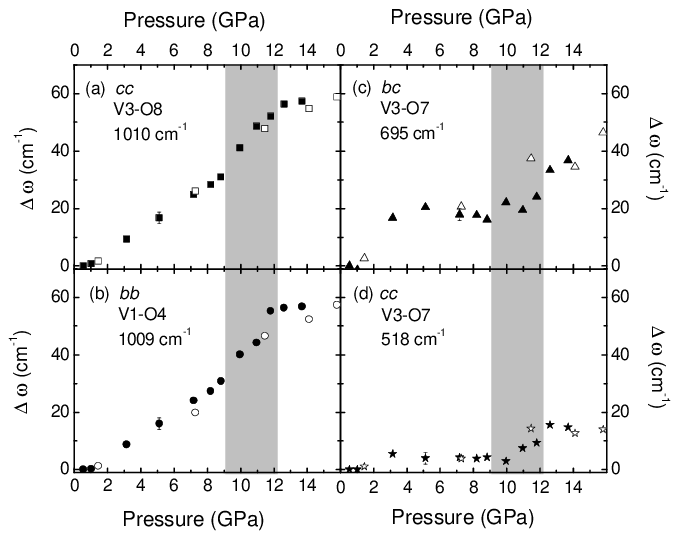,width=90mm,clip=}}
\caption{Pressure dependence of the frequencies of several Raman
modes of $\beta$-Na$_{0.33}$V$_2$O$_5$ at room temperature;
full and open symbols denote results of two experimental runs. (a)
V3-O8 bond stretching mode at 1010~cm$^{-1}$. (b) V1-O4 stretching mode at
1009~cm$^{-1}$. (c) V3-O7 stretching mode along the $b$ axis at 695~cm$^{-1}$
(d) V3-O7 stretching mode along the $c$ axis at 518~cm$^{-1}$. The grey bars
indicate the pressure range (9 - 12~GPa) with the most
pronounced pressure-induced changes.
\label{phononfreq}}
\end{figure}

A closer inspection of the spectra, however, reveals interesting
pressure-induced changes:
Almost all modes harden with increasing pressure, and the pressure-induced
shift depends on the pressure range. In Fig.\ \ref{phononfreq}
the frequency shifts of the Raman modes with the most pronounced changes
are shown. The frequency positions were obtained by fitting with
Lorentzian functions. Below we describe the pressure-induced changes
in more detail. All these changes are reversible upon pressure release.

For \emph{cc} polarization [see Fig.\ \ref{pressall}(a)] the V3-O8 stretching mode
located in the high-frequency range shifts
linearly in the pressure range $\leq$9~GPa. The linear
pressure coefficient was obtained by fitting the peak positions
with the function $\omega(P)=A+B*P$, where $P$ is the applied pressure.
The so-obtained linear pressure coefficient $B$ for this mode and the other
modes is included in Table\ \ref{Rphonon}. Above 9~GPa the V3-O8 stretching mode
hardens strongly, and a saturation sets in above approximately
12~GPa [Fig.\ \ref{phononfreq}(a)].
In comparison, the V3-O7 stretching mode shows almost no pressure-induced
frequency change up to approx.\ 10~GPa [Fig.\ \ref{phononfreq}(d)].
Between 10 and 12~GPa its frequency increases, and above 12~GPa the mode remains 
approximately constant.
Furthermore, above 9~GPa a new Raman mode appears at a frequency of
230~cm$^{-1}$, which hardens with increasing pressure. The broad
Raman peak at 702~cm$^{-1}$ loses intensity with increasing pressure and
evolves into a narrower peak, located at 758~cm$^{-1}$ for the highest
applied pressure.

For \emph{bb} polarization [see Fig.\ \ref{pressall}(b)] the
strongest modes at 266 and 440~cm$^{-1}$ are not much affected by
the pressure application regarding their frequency position: For
the pressure range below 9~GPa they shift only by a few
wavenumbers, and above 9 GPa their positions are approximately
pressure-independent. The frequency of the V1-O4 stretching mode
increases linearly with increasing pressure [see Fig.\
\ref{phononfreq}(b)] and remains constant above 12~GPa. It is
interesting to note that the mode at 266~cm$^{-1}$ appreciably
loses intensity above 9 GPa. A further change induced at around 9~GPa 
is the appearance of a Raman mode at around 230~cm$^{-1}$,
whose intensity increases with increasing pressure. 
In addition, two new modes (located at 500 and 550~cm$^{-1}$) appear 
above 5~GPa, shifting to higher frequencies with increasing pressure.

For \emph{bc} polarization  the V3-O7 stretching mode shows the
strongest pressure dependence [see Figs.\ \ref{pressall}(c) and
\ref{phononfreq}(c)]: After an initial shift, the mode is
approximately pressure-independent up to 12 GPa, and above 12 GPa
it hardens. Furthermore, above 9 GPa a weak, pressure-independent Raman mode
appears at around 330~cm$^{-1}$.

\begin{figure}[t]
\centerline{\psfig{file=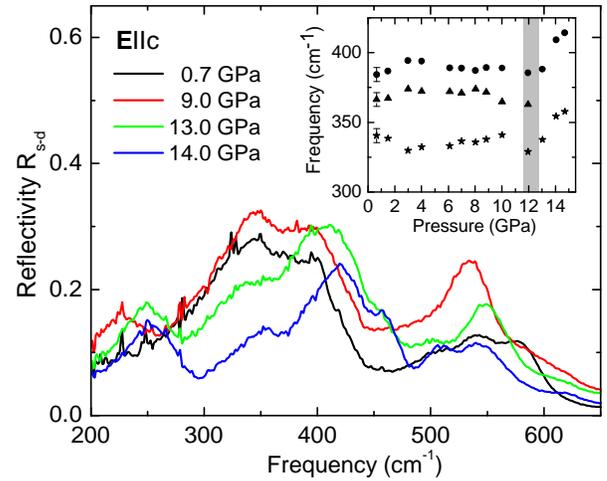,width=80mm,clip=}}
\caption{(Color online) Far-infrared {\bf E}$\parallel$$c$ reflectivity spectra 
$R_{s-d}$ of $\beta$-Na$_{0.33}$V$_2$O$_5$ at room temperature for selected pressures.
Inset: Frequency positions as a function of pressure for several phonon
modes, obtained by fitting the reflectivity spectra with the Drude-Lorentz model;
the grey bar indicates the pressure range with the most-pronounced pressure-induced
changes.
\label{far-infrared}}
\end{figure}

In summary, for all the different studied polarizations the most pronounced
changes in the Raman spectra are found in the pressure range between 9 and
12~GPa. This is in good agreement
with our pressure-dependent far-infrared reflectivity measurements:
In Fig.\ \ref{far-infrared} the room-temperature reflectivity spectra $R_{s-d}$
of $\beta$-Na$_{0.33}$V$_2$O$_5$ for the polarization {\bf E}$\parallel$$c$ are
shown for selected pressures. We obtained the frequency positions of the phonon modes
by fitting the spectra with the Drude-Lorentz model
combined with the normal-incidence Fresnel equation, taking into account the
known refractive index of diamond. Several modes show marked changes at around 
12~GPa, which is illustrated in the inset of Fig.\ \ref{far-infrared}:
Here we show the pressure dependence of the frequency positions of the modes
in the range 300 - 400~cm$^{-1}$, which can be assigned to the polyhedral bending 
modes. \cite{Thirunavukkuarasu06} 
While the frequency positions of these modes are almost pressure independent below 12~GPa,
they significantly shift to higher energies above this pressure value. In addition, 
the oscillator strengths of the modes change at $\approx$12~GPa.
At around the same pressure the changes in the Raman-active modes are observed, suggesting
a conjoint interpretation of the effects.

\begin{table*}[h!]
\caption{Room-temperature Raman modes of
$\beta$-Na$_{0.33}$V$_2$O$_5$ with their linear pressure coefficients, obtained
by fitting their frequency position with the expression $\omega(P)=A+B*P$, and their assignment.
Modes marked with an asterisk (dagger) appear (disappear) in the pressure range 9 - 12~GPa.}
\label{Rphonon}
\begin{center}
\begin{ruledtabular}
\begin{tabular}{|l|l|c|c|c|c|}
    geometry & frequency $\omega$ & B for P$\leq$9GPa & B for 9GPa$<$P$\leq$12GPa & B for P$>$12GPa  & assignment\\
    & (cm$^{-1}$)&(cm$^{-1}$/GPa) & (cm$^{-1}$/GPa)&(cm$^{-1}$/GPa) & \\
    \hline
       &230* &  & 6.5& 6.5& \\
       $cc$ &298$^\dag$ & 2.5 &  &  & \\
       &518 & 0.3 & 4.5 & -0.8 & V3-O7 $c$ axis stretching\\
       &1010 & 3.8 & 6.0 & 0.6 & V3-O8 stretching\\

    \hline
     &230*&  &  &  & \\
     &266 & 1.4 & 0.2  & 0.2 & \\
     $bb$ &331$^\dag$ & 2.8 &  & & \\
      &390 & 0.1 & 0.1 & 0.1 &\\
      &440 & 1.7 & -1.9 & 1.4 & V2-O1-V2 bending\\
      &500* & &&&\\
      & 550*& &&&\\
      &1009 & 3.7 & 7.7 & 0.7 & V1-O4 stretching\\

     \hline
     &138*& &&&\\
      &154 & 1.2 & 1.2 & 1.2 & \\
      &165 & &  & & \\
     $bc$ &256  & 0.1 & 0.1 & 0.1 &\\
      &275$^\dag$ & 1.8 & &  & \\
      &330*&&&&\\
      &695 & 0.5 & 1.9 & -0.6 & V3-O7 $b$ axis stretching\\

      \hline
  \end{tabular}
  \end{ruledtabular}
  \end{center}
\end{table*}

\section{Discussion} \label{sectiondiscussion}

Earlier pressure-dependent mid-infrared data of $\beta$-Na$_{0.33}$V$_2$O$_5$
found new excitations induced at around 12~GPa, which were interpreted in terms
of a redistribution of charge among the different V sites.\cite{Kuntscher05}
Based on these data, a pressure-induced structural phase transition or amorphization
of the sample appeared unlikely. The occurrence of a structural phase transition, 
however, could not be ruled out.
According to our results, most of the Raman and far-infrared modes are present up
to the highest applied pressure. Therefore, we can rule out an amorphization of 
the sample between 10 and 12~GPa. Furthermore, our results show that the structural 
units, i.e., the V-O polyhedra, remain intact up to the highest applied pressure. 

In the pressure range 10 - 12~GPa several Raman-active modes show a significant change in their
pressure-dependent frequency shifts. In general, the force constant
and hence the frequency of a Raman mode are affected by the amount of charge
on the ions involved in the vibrations.\cite{Bacsa00,Popovic02}
Accordingly, we interpret the changes observed in our Raman data in terms of a transfer
of charge between the different V sites, setting in at 9~GPa and being
completed at around 12~GPa. It is interesting to note, that the frequency of the
V2-O1-V2 bending mode located at 440~cm$^{-1}$ in {\it bb} polarization
is hardly affected by the pressure application. Therefore, we speculate that the
charge transfer occurs mainly among the V1 and V3 sites.

According to the above interpretation, the amount of charge located on the different 
structural entities [(V1)O$_6$ and (V2)O$_6$ octahedra, (V3)O$_5$ polyhedra] is changed 
between 9 and 12~GPa, which should influence the frequencies of the related infrared 
modes (like polyhedral bending and stretching modes). Indeed, this is demonstrated 
by the results of our far-infrared reflectivity data. Besides the frequency positions
of the modes, their oscillator strengths are altered at around 12~PGa, which 
is also consistent with a pressure-induced rearrangement of the charges.

A further issue of interest concerns the relevance of
electron-phonon coupling and possible formation of polaronic
quasiparticles in $\beta$-Na$_{0.33}$V$_2$O$_5$. The pronounced
mid-infrared band observed by reflectivity measurements was
attributed to polaronic excitations.\cite{Presura03} Its
pressure-induced redshift up to 12~GPa confirms this
interpretation.\cite{Kuntscher05,Kuntscher06,Kuntscher062}
Additional support of the importance of electron-phonon coupling
can be inferred from the presence of the broad Raman mode in the
{\it cc} Raman spectrum at around 700~cm$^{-1}$ and its pressure
dependence: Such a broad mode was also found in the Raman spectrum
of the closely related $\alpha$'-NaV$_2$O$_5$. Here, its origin
was related to the coupling of a phonon mode to an electronic
state, among other possible scenarios.\cite{Fischer99} A similar
mechanism based on electron-phonon coupling might also lead to the
broad Raman mode in the present material. The mode is clearly
visible in the Raman spectrum of $\beta$-Na$_{0.33}$V$_2$O$_5$ at
ambient pressure, but weakens with increasing pressure and seems
to have disappeared completely above $\approx$10~GPa. In the same
pressure range (10 - 12~GPa) the presumable polaronic band in our
mid-infrared reflectivity
spectra\cite{Kuntscher05,Kuntscher06,Kuntscher062} changes its
character. These findings suggest that electron-phonon coupling
is indeed important in $\beta$-Na$_{0.33}$V$_2$O$_5$, at least up
to $\approx$12~GPa.

\section{Summary}
Our polarization-dependent Raman and far-infrared reflectivity spectra of
$\beta$-Na$_{0.33}$V$_2$O$_5$ single crystals under pressure show significant 
changes in the phonon modes for pressures 9 - 12~GPa. Since most of the 
spectral features are present up to the highest applied pressure, a
pressure-induced amorphization of the sample between 10 and 12~GPa
can be ruled out. Furthermore, the structural units, i.e., the V-O 
polyhedra, remain intact up to the highest pressure. 
The observed pressure-induced changes in the optical properties can be related 
to a pressure-induced transfer of charge among the different V sites.
The presence of the broad Raman mode in {\it cc} polarization below $\approx$12~GPa
suggests the relevance of electron-phonon coupling in the low-pressure range.

\acknowledgements
We thank G. Untereiner for technical assistance. We acknowledge the ANKA
Angstr\"omquelle Karlsruhe for the provision of beamtime and we would like to thank
D. Moss, Y.-L. Mathis, B. Gasharova, and M. S\"upfle for assistance using beamline ANKA-IR.
Financial support by the DAAD and the Deutsche Forschungsgemeinschaft through 
the Emmy Noether-program and SFB 484 is gratefully acknowledged.


\begin{thebibliography}{99}
\item[$^*$] Email: christine.kuntscher@physik.uni-augsburg.de

\bibitem{Yamauchi02}
T. Yamauchi, Y. Ueda, and N. M\^ori, Phys.\ Rev.\ Lett.\ {\bf 89},
057002 (2002).

\bibitem{Itoh06}
M. Itoh, I. Yamauchi, T. Kozuka, T. Suzuki, T. Yamauchi, J.-I.
Yamaura, and Y. Ueda, Phys.\ Rev.\ B {\bf 74}, 054434 (2006).

\bibitem{Wadsley55}
A.D. Wadsley\ Acta Cryst.\ {\bf 8}, 695 (1955)

\bibitem{Doublet05}
 M.-L. Doublet, and M.-B. Lepetit, Phys.\ Rev.\ B {\bf 71}, 075119 (2005).

\bibitem{Yamada99}
H. Yamada and Y. Ueda, J.\ Phys.\ Soc.\ Jpn.\ \textbf{68}, 2735
(1999).

\bibitem{Heinrich04}
M. Heinrich, H.-A. Krug von Nidda, R. M. Eremina, A. Loidl, Ch.
Helbig, G. Obermeier and S. Horn, Phys.\ Rev.\ Lett.\ \textbf{93},
116402 (2004).

\bibitem{Ueda01}
Y. Ueda, H. Yamada, M. Isobe, and T. Yamauchi, J.\ Alloys\ Compd.\
\textbf{317-318}, 109 (2001).

\bibitem{Presura03}
C. Presura, M. Popinciuc, P. H. M. van Loosdrecht, D. van der
Marel, M. Mostovoy, T. Yamauchi and Y. Ueda, Phys.\ Rev.\ Lett.
\textbf{90}, 2, 026402 (2003).

\bibitem{Kuntscher05}
C. A. Kuntscher, S. Frank, I. Loa, K. Syassen, T. Yamauchi, and Y.
Ueda, Phys.\ Rev.\ B {\bf 71}, 220502(R) (2005).

\bibitem{Kuntscher06}
C. A. Kuntscher, S. Frank, I. Loa, K. Syassen, T. Yamauchi, and Y.
Ueda, Physica B {\bf 378-380}, 896-897 (2006).

\bibitem{Kuntscher062}
C. A. Kuntscher, S. Frank, I. Loa, K. Syassen, F. Lichtenberg, T.
Yamauchi, and Y. Ueda, Infrared Physics \& Technology {\bf 49}, 88
(2006).

\bibitem{Mao86}
H. K Mao, J. Xu, and P.M. Bell, J.Geophys. Res [Atmos.]{\bf 91},
4673 (1986).

\bibitem{Popovic03}
Z. V. Konstantinovi\ifmmode \acute{c}\else \'{c}\fi{}, M. J.
Popovi\ifmmode \acute{c}\else \'{c}\fi{}, V. V. Moshchalkov, M.
Isobe and Y. Ueda, J.\ Phys.\ Cond.\ Matter\ {\bf 15}, L139-L145
(2003).

\bibitem{Isobe00}
 M. Isobe and Y. Ueda, Mol.\ Cryst.\ Liq.\ Cryst.\ {\bf 341}, 1075
(2000).

\bibitem{Yamura02}
J. I. Yamura, M. Isobe, H. Yamada, T.  Yamauchi, and Y. Ueda, J.\
Phys.\ Chem.\ Solids\ {\bf 63}, 957 (2002).

\bibitem{Isobe98}
M. Isobe and Y. Ueda, J. Phys.\ Soc.\ Jpn.\ {\bf 65}, 1178 (1996);
K. Ohwada, Y. Fujii, Y. Katsuki, J. Muraoka, H. Nakao, Y.
Murakami, H. Sawa, E. Ninomiya, M. Isobe, and Y. Ueda, Phys.\
Rev.\ Lett.\ {\bf 94}, 106401 (2005) and references therein.

\bibitem{Konstantinovic02}
M. J. Konstantinovi\ifmmode \acute{c}\else \'{c}\fi{}, Z. V.
Popovi\ifmmode \acute{c}\else \'{c}\fi{},
 V. V. Moshchalkov, C. Presura, R. Gaji\ifmmode \acute{c}\else \'{c}\fi{}, M. Isobe, and Y. Ueda,
 Phys.\ Rev.\ B\ {\bf 65}, 245103 (2002).


\bibitem{Bacsa00}
 W. S. Bacsa, R. Lewandowska, A. Zwick, and P. Millet, J.\ Phys.\ Rev.\ B\ {\bf 61}, R14885 (2000).

\bibitem{Golubchik97}
S. A. Golubchik, M. Isobe, A. N. Ivlev, B. N. Mavrin, M. n.
Popova, A. B. Sushkov, Y. Ueda, and A. N. Vesil'ev, J.\ Phys.\
Soc.\ Jpn.\ \textbf{66}, 4042 (1997).

\bibitem{Loa99}
I. Loa, U. Schwarz, M. Hanfland, R. K. Kremer, and K. Syassen,
phys.\ stat.\ sol.\ (b) {\bf 215}, 709 (1999).

\bibitem{Konstantinovic99}
M. J. Konstantinovi\ifmmode \acute{c}\else \'{c}\fi{}, K. Ladavac,
A. Beli\ifmmode \acute{c}\else \'{c}\fi{}, A. N. Vesil'ev, M.
Isobe, and Y. Ueda, J. Phys.: Condens.\ Matter {\bf 11}, 2103
(1999).

\bibitem{Fischer99}
 M. Fischer, P. Lemmens, G. Els, G. G\"untherodt, E.Y. Sherman, E. Morr\'e, C. Geibel,
 and F. Steglich J.\ Phys.\ Rev.\ B\ {\bf 60}, 7284 (1999).

\bibitem{Popovic02}
 Z. V. Popovi\ifmmode \acute{c}\else \'{c}\fi{}, M. J.
Konstantinovi\ifmmode \acute{c}\else \'{c}\fi{}, R. Gaji\ifmmode
\acute{c}\else \'{c}\fi{}, V. N. Popov, M. Isobe, Y. Ueda, and V.
V. Moshchalkov,  Phys.\ Rev.\ B {\bf 65}, 184303 (2002).

\bibitem{Thirunavukkuarasu06}
K. Thirunavukkuarasu, F. Lichtenberg, and C. A. Kuntscher, 
J. Phys.: Condens.\ Matter {\bf 18}, 9173 (2006). 
 


\end{thebibliography}
\end{document}